\documentclass[%
reprint,
superscriptaddress,
showpacs,
preprintnumbers,
 bibnotes,
 amsmath,amssymb,
]{revtex4-1}

\usepackage{graphicx}% Include figure files
\usepackage{dcolumn}% Align table columns on decimal point
\usepackage{bm}% bold math
\usepackage{color}
\begin{document}

%\preprint{}

\title{Nematic magnetoelastic effect contrasted between Ba(Fe$_{1-x}$Co$_{x}$)$_2$As$_2$ and FeSe}

\author{Yuwen~Hu}
\altaffiliation[Present address: ]{Department of Physics, Princeton University, Princeton, NJ 08544, USA}
\affiliation{International Center for Quantum Materials, School of Physics, Peking University, Beijing 100871, China}
\author{Xiao~Ren}
\affiliation{International Center for Quantum Materials, School of Physics, Peking University, Beijing 100871, China}
\author{Rui~Zhang}
\affiliation{Department of Physics and Astronomy, Rice University, Houston, Texas 77005, USA}
\author{Huiqian~Luo}
\affiliation{Beijing National Laboratory for Condensed Matter Physics, Institute of Physics, Chinese Academy of Sciences, Beijing 100190, China}
\author{Shigeru~Kasahara}
\affiliation{Department of Physics, Kyoto University, Kyoto 606-8502, Japan}
\author{Tatsuya~Watashige}
\affiliation{Department of Physics, Kyoto University, Kyoto 606-8502, Japan}
\author{Takasada~Shibauchi}
\affiliation{Department of Advanced Materials Science, University of Tokyo, Chiba 277-8561 Japan}
\author{Pengcheng Dai}
\affiliation{Department of Physics and Astronomy, Rice University, Houston, Texas 77005, USA}
\author{Yan~Zhang}
\affiliation{International Center for Quantum Materials, School of Physics, Peking University, Beijing 100871, China}
\affiliation{Collaborative Innovation Center of Quantum Matter, Beijing 100871, China}
\author{Yuji~Matsuda}
\affiliation{Department of Physics, Kyoto University, Kyoto 606-8502, Japan}
\author{Yuan~Li}
\email{yuan.li@pku.edu.cn}
\affiliation{International Center for Quantum Materials, School of Physics, Peking University, Beijing 100871, China}
\affiliation{Collaborative Innovation Center of Quantum Matter, Beijing 100871, China}

\begin{abstract}
To elucidate the origin of nematic order in Fe-based superconductors, we report a Raman scattering study of lattice dynamics, which quantify the extent of $C_4$-symmetry breaking, in BaFe$_2$As$_2$ and FeSe. FeSe possesses a nematic ordering temperature $T_\mathrm{s}$ and orbital-related band-energy split below $T_\mathrm{s}$ that are similar to those in BaFe$_2$As$_2$, but unlike BaFe$_2$As$_2$ it has no long-range magnetic order. We find that the $E_g$ phonon-energy split in FeSe sets in only well below $T_\mathrm{s}$, and its saturated value is substantially smaller than that in BaFe$_2$As$_2$. Together with reported results for the Ba(Fe$_{1-x}$Co$_{x}$)$_2$As$_2$ family, the data suggest that magnetism exerts a major influence on the lattice.
\end{abstract}


\maketitle

In copper- and iron-based high-temperature superconductors, as well as in heavy-fermion and organic superconductors, the superconducting phase is commonly found in close proximity to an antiferromagnetic phase. Not only does this important commonality suggest that the mechanism for unconventional superconductivity builds upon electronic correlations that give rise to the magnetism \cite{MoriyaRepProgPhys2003,LeeRMP2006,MonthouxNature2007,UemuraNatMater2009,NormanScience2011,WangScience2011,ScalapinoRMP2012,DaiRMP2015}, but it also implies that intriguing ``intertwined phases'', which have been a subject of intense study \cite{FradkinNatPhys2012,KeimerNature2015}, may arise from the same electronic correlations \cite{DavisPNAS2013}. In the Fe-based superconductors \cite{BasovNatPhys2011}, the most prominent intertwined phase is the so-called nematic phase \cite{FisherRepProgPhys2011,KasaharaNature2012,FradkinAnnuRev2010}, in which the discrete $C_4$ rotational symmetry is broken but the lattice translational symmetry is not. Because electronic properties exhibit pronounced $C_2$ (rather than $C_4$) symmetry in the nematic phase while the crystal structure is only weakly orthorhombic \cite{ChuScience2010,ChuangScience2010,LiPreprint2015}, there has been general consensus that the nematic phase is electronically driven \cite{FernandesNatPhys2014}. The possible existence of a nematic quantum critical point has been intensively explored in this context \cite{HashimotoScience2012,ShibauchiAnnuRev2014}, as it might explain some of the most unusual properties of these materials including the superconductivity itself.

Consistent with the notion that all essential intertwined phases in unconventional superconductors arise from a common magnetic origin \cite{DavisPNAS2013}, the tendency towards formation of stripe antiferromagnetic order in the Fe-based superconductors is considered a likely driving force for the nematic order. Such theoretical ideas have been explored in contexts both with \cite{FangPRB2008,XuPRB2008,FernandesPRB2012} and without \cite{Wang2015,YuPRL2015,Glasbrenner2015,ChubukovPRB2015} stripe antiferromagnetic order as the system's low-temperature ground state. The latter theories are motivated by the case of bulk FeSe \cite{McQueenPRL2009}, which exhibits a nematic transition at $T_\mathrm{s} \approx 90$ K but no long-range magnetic order down to the lowest temperature.

However, photoemission studies \cite{YiPNAS2011,ZhangPRB2012,YiNJP2012,ShimojimaPRB2014,NakayamaPRL2014} have revealed below $T_\mathrm{s}$ a dramatic electronic reconstruction, which leads to an uneven occupation of the Fe $d_{xz}$ and $d_{yz}$ orbitals. When the magnetic ordering temperature $T_\mathrm{mag}$ is well below $T_\mathrm{s}$, the reconstruction has been reported to be seen already above $T_\mathrm{mag}$ \cite{ZhangPRB2012,YiNJP2012}, although the effect of detwinning uniaxial pressure on $T_\mathrm{mag}$ \cite{DhitalPRL2012,DhitalPRB2014} remains yet be considered. The electronic reconstruction in the pnictides improves the quality of Fermi-surface nesting \cite{YiNJP2012}, which can in turn help stabilize the stripe antiferromagnetic order. Together with the absence of long-range magnetic order and of anomaly in the low-energy spin fluctuations near $T_\mathrm{s}$ \cite{BaekNatMater2015,BoehmerPRL2015}, yet similarly pronounced electronic reconstruction in FeSe \cite{ShimojimaPRB2014,NakayamaPRL2014,WatsonPRB2015FeSe,KasaharaPNAS2014,SuzukiPreprint2015} as in other systems, these results support the alternative scenario that the nematic order is driven by orbital interactions \cite{LeePRL2009,KruegerPRB2009,BasconesPRL2010,LvPRB2010,ChenPRB2010} or by a related Pomeranchuk instability \cite{ZhaiPRB2009}. To what extent some of the most recent results can be thought of as refuting spin-driven and/or ferro-orbital nematic order is currently under heated debate \cite{WatsonPRB2015FeSeS,ZhangPRB2015,ZhangDHLu2015,MukherjeePRL2015}.

To experimentally determine whether the nematic order is spin- or orbital-driven, in principle one would need to measure the susceptibility of spin-correlation anisotropy to orbital polarization, or vice versa, much in the fashion of what has been achieved between the electronic and lattice degrees of freedom \cite{ChuScience2012}, but this is obviously difficult. Here we take an alternative approach by using lattice dynamics to detect the ``strength'' of nematicity in BaFe$_2$As$_2$ and FeSe. Since the lattice is linearly coupled to the electronic nematicity \cite{BoehmerReview2015}, and because the lattice (as we will show) and orbital-related \cite{ShimojimaPRB2014,YiPNAS2011} characteristic energies are respectively similar between the two systems, our measurement can determine how spin structures substantiate the nematic order. We find that the lattice-dynamics signature of $C_4$-symmetry breaking in FeSe only sets in below $T^*\sim 65$ K rather than immediately below $T_\mathrm{s}$, and that its saturated value is substantially smaller than that in BaFe$_2$As$_2$. Our results suggest that spin supersedes orbital in causing nematic lattice deformations.

BaFe$_2$As$_2$ is the parent compound of the ``122'' family Fe-based superconductors, exhibiting an orthorhombic stripe antiferromagnetic phase below $T_\mathrm{mag} \approx T_\mathrm{s} = 138$ K \cite{KimPRB2011}. FeSe is structurally the simplest Fe-based superconductor with an orthorhombic structural transition at $T_\mathrm{s} \approx 90$ K but no long-range magnetic order \cite{McQueenPRL2009}. At high temperatures, BaFe$_2$As$_2$ and FeSe belong to the $I4/mmm$ and $P4/nmm$ space groups, respectively, and the Fe and As/Se atoms contribute two two-fold degenerate $E_g$ phonon modes at the Brillouin zone center. When the $C_4$ rotational symmetry is lowered into $C_2$ in the nematic phase, each of the $E_g$ modes splits into $B_{2g}$ and $B_{3g}$ modes that are of slightly different energies. Since all these phonons are Raman-active, we can utilize the high energy resolution and sensitivity of Raman scattering to detect the energy split, which provides information about the $ab$-anisotropy of the lattice ``spring constants'' arising from the spin and/or orbital interactions.

\begin{figure}
\includegraphics[width=2.9in]{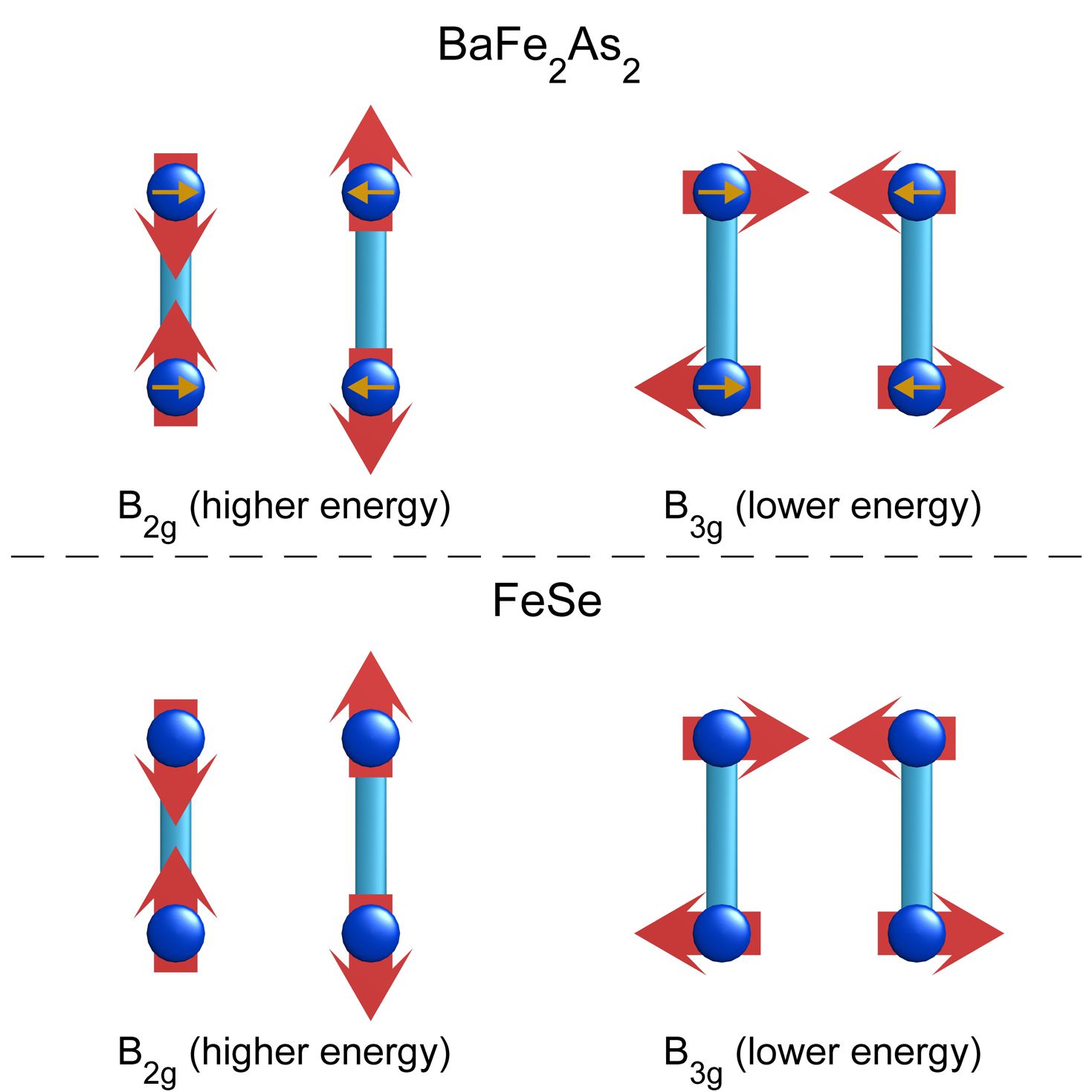}
\caption{Thick arrows indicate displacement of Fe atoms in the $B_{2g}$ and $B_{3g}$ phonon modes in BaFe$_2$As$_2$ and FeSe. As/Se atoms are omitted for clarity. The vertical Fe-Fe bonds are highlighted as being more rigid than the horizontal ones. }
\label{Fig1}
\end{figure}

We performed our variable-temperature Raman scattering experiment in a confocal backscattering geometry, using a Horiba Jobin Yvon LabRAM HR Evolution spectrometer equipped with 1800 gr/mm gratings and a liquid-nitrogen-cooled CCD detector. Long-wavelength $\lambda$= 785 nm and 633 nm lasers were used as excitations to achieve high energy resolution. We kept our laser power low ($\sim$1 mW) to reduce heating, which led to very long exposure time ($>10$ hours per spectrum) in order to obtain satisfactory statistics in the photon counts. Samples were kept in a cryostat under better than $5\times10^{-8}$ Torr vacuum to ensure surface stability over the entire measurements. High-quality single crystals of BaFe$_2$As$_2$ and FeSe were grown by self-flux and chemical vapor transport methods, respectively. The Raman measurements were performed on surfaces that are perpendicular to the easy-cleavage $ab$-plane, which allowed us to use perpendicular linear polarizations of incoming and scattered photons to detect the $E_g$, $B_{2g}$, and $B_{3g}$ phonons. Such sample surfaces were prepared by cleaving the crystals after freezing in liquid nitrogen.

Figure~\ref{Fig1} illustrates the vibrational patterns of Fe atoms in $B_{2g}$ and $B_{3g}$ modes that derive from the same $E_g$ mode in the high-temperature phase. Because of the uneven $d_{xz}$ and $d_{yz}$ orbital occupation, bonds along one of the Fe-Fe directions is expected to be stronger, and atomic vibrations along that direction are expected to occur at slightly higher frequency (or energy). The difference between BaFe$_2$As$_2$ and FeSe is that the former also exhibits a stripe antiferromagnetic order, which is expected to further influence the lattice dynamics via magnetoelastic coupling \cite{ChauvierePRB2009}. The question is how large such effects are compared to the influence of the orbital and/or Fermi-surface anisotropy in the nematic phase. Importantly, photoemission experiments have found comparable magnitudes of orbital-related band-energy split in the two systems \cite{YiPNAS2011,ShimojimaPRB2014,NakayamaPRL2014}, so any substantial difference we identify has to arise from the difference in the magnetism.

\begin{figure}
\includegraphics[width=2.9in]{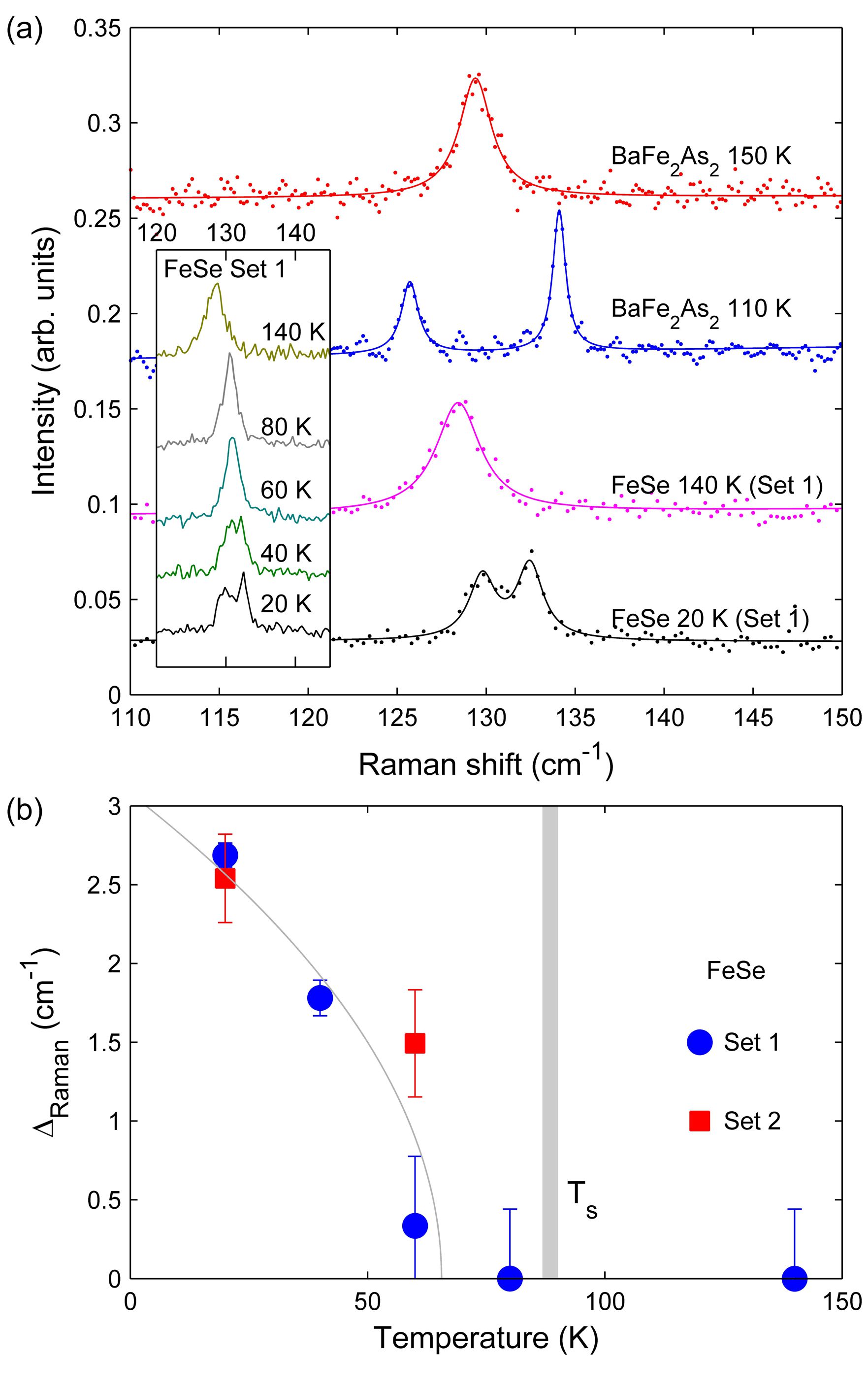}
\caption{(a) Raman spectra measured on BaFe$_2$As$_2$ and FeSe single crystals. The inset shows temperature dependence of the spectrum of FeSe near 130 cm$^{-1}$, vertically offset for clarity. (b) Temperature dependence of phonon-energy split in FeSe. The two data sets were obtained using different laser powers, from which we conclude that laser heating is not an issue in our experiment.}
\label{Fig2}
\end{figure}

We present our key result in Fig.~\ref{Fig2}. At high temperatures, $T=150$ K $>T_\mathrm{s}$ in BaFe$_2$As$_2$ and $T=140$ K $>T_\mathrm{s}$ in FeSe, the $E_g$ phonon peaks of both systems are observed at very similar energies. This shows that the two systems possess similar lattice dynamics in the tetragonal phase, which is not an unexpected result given the similar atomic masses of As and Se and the structural similarity between the FeAs and FeSe layers. As we have recently reported \cite{Ren2015}, at $T=110$ K $<T_\mathrm{s}$ in BaFe$_2$As$_2$, the $E_g$ peak splits into $B_{2g}$ and $B_{3g}$ peaks that differ in energy by 9.4 cm$^{-1}$, consistent with a previous report \cite{ChauvierePRB2009}. In contrast, although a splitting of the $E_g$ peak is also observed at $T=20$ K $\ll T_\mathrm{s}$ in FeSe, the $B_{2g}$ and $B_{3g}$ peaks only differ in energy by 2.6 cm$^{-1}$. We attribute the much smaller energy split in FeSe to the lack of magnetic order as discussed above.

A further unexpected observation is that, unlike in BaFe$_2$As$_2$, where the phonon-energy split rapidly increases below $T_\mathrm{s}$ and reaches its saturated value about 30 K below $T_\mathrm{s}$ \cite{ChauvierePRB2009}, the split in FeSe is not observed immediately below $T_\mathrm{s}$. Instead, it only develops below $T^*\approx65$ K, as shown in the inset of Fig.~\ref{Fig2}(a) and Fig.~\ref{Fig2}(b). This value of $T^*$ is consistent with the temperature below which the spin-lattice relaxation rate is found to increase \cite{BaekNatMater2015,BoehmerPRL2015}. Thus the phonon-energy split in FeSe, albeit small and in the absence of static magnetic order, might nevertheless be caused by low-energy spin fluctuations which are presumably nematic in nature.

\begin{figure}
\includegraphics[width=2.9in]{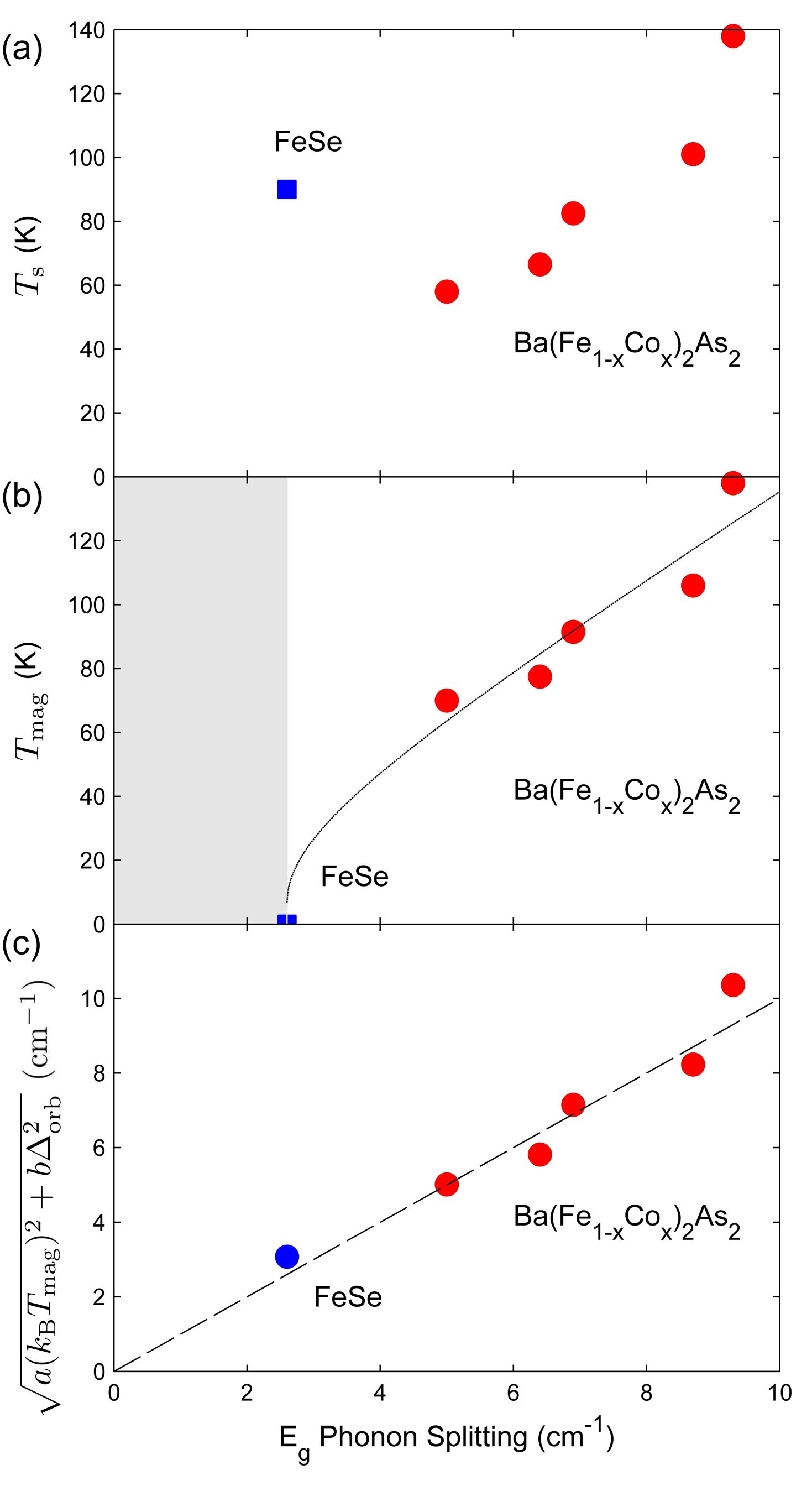}
\caption{Structural phase transition temperature (a), magnetic ordering temperature (b), and an empirical combination (see text) of magnetic and orbital energies (c) plotted versus $E_g$ phonon-energy split for both FeSe and Ba(Fe$_{1-x}$Co$_x$)$_2$As$_2$.}
\label{Fig3}
\end{figure}

In the Ba(Fe$_{1-x}$Co$_x$)$_2$As$_2$ family, the phonon-energy split is found to decrease with increasing Co doping \cite{ChauvierePRB2009}, which simultaneously suppresses the tetragonal-to-orthorhombic structural transition temperature $T_\mathrm{s}$, the stripe antiferromagnetic ordering temperature $T_\mathrm{mag}$ \cite{NiPRB2010}, the orbital-related band-energy split $\Delta_\mathrm{orb}$ \cite{YiPNAS2011}, and the transport anisotropy \cite{ChuScience2010}. It is therefore difficult to decipher the relationship among these quantities by studying this material family alone. To this end, we have attempted to empirically relate the phonon-energy split to the magnetic and orbital characteristic energies, accommodating both Ba(Fe$_{1-x}$Co$_x$)$_2$As$_2$ and FeSe.

Our results are presented in Fig.~\ref{Fig3}. First of all, we find that the phonon-energy split is not simply related to the structural transition temperature [Fig.~\ref{Fig3}(a)]. Despite its likely connection to the magnetism as discussed in the preceding paragraphs, the split is not simply linearly related to $T_\mathrm{mag}$ either, as in that case the split in FeSe would be expected to be nearly zero [Fig.~\ref{Fig3}(b)]. The split in FeSe appears to be bounded from below by another mechanism, which we assume here to be orbital interactions. By considering the reported values of phonon-energy split $\Delta_\mathrm{Raman}$ \cite{ChauvierePRB2009} and $\Delta_\mathrm{orb}$ \cite{YiPNAS2011} as functions of Co concentration $x$, which has a one-to-one correspondence to $T_\mathrm{mag}$ in Ba(Fe$_{1-x}$Co$_x$)$_2$As$_2$ \cite{ChuScience2010,NiPRB2010}, we find that the empirical formula $\Delta_\mathrm{Raman}=\sqrt{a (k_\mathrm{B} T_\mathrm{mag})^2+ b \Delta_\mathrm{orb}^2}$, where $a$ and $b$ are dimensionless parameters, describes all the data very well [Fig.~\ref{Fig3}(c)]. The underlying assumption for this formula is that the spin-related energy $k_\mathrm{B} T_\mathrm{mag}$ and the orbital-related energy $\Delta_\mathrm{orb}$ influence the lattice dynamics in an uncorrelated fashion. We find that $a=1.0 \times 10^{-2}$, which is much greater than $b=5.8 \times 10^{-5}$, \textit{i.e.}, the spin correlations exert a much stronger influence on the lattice than the orbital structure, as expected from the fact $\Delta_\mathrm{orb} = 62$ meV and 50 meV in BaFe$_2$As$_2$ and FeSe \cite{YiPNAS2011,ShimojimaPRB2014}, respectively, yet their phonon-energy splits differ by over a factor of three.

\begin{figure}
\includegraphics[width=2.9in]{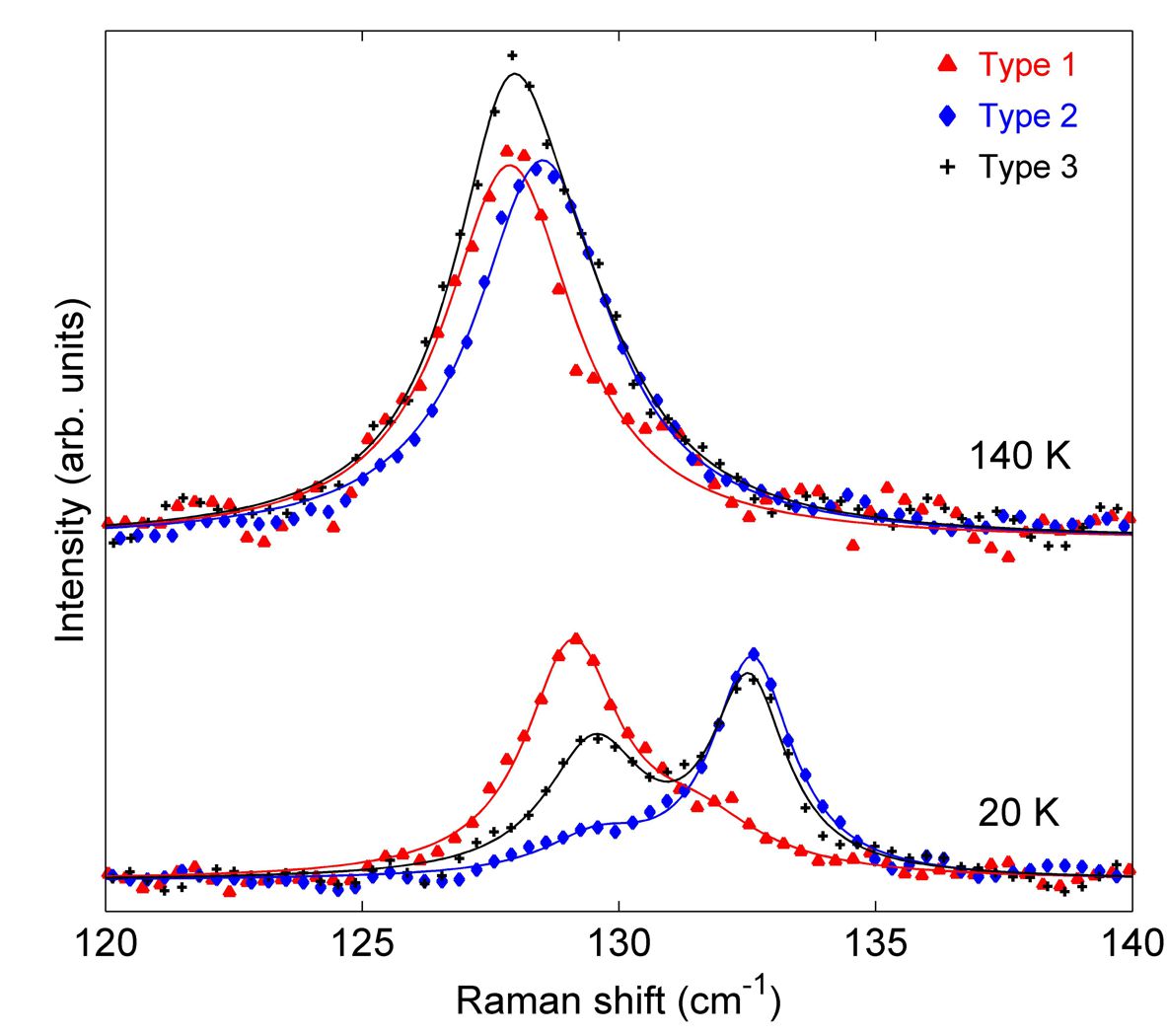}
\caption{Raman spectra measured on three representative surface spots of FeSe with different local stress.}
\label{Fig4}
\end{figure}

In the above analysis, we have used $\Delta_\mathrm{orb}$ determined from photoemission experiments, some of which were performed on samples detwinned by uniaxial stress. Since we did not use a detwinned sample here, and because magnetic and transport properties are sensitive to uniaxial pressure especially near $T_\mathrm{s}$, it is important to check the possible influence of local stress on our result. Indeed, we have identified three types of surface spots on our FeSe sample, as shown in Fig.~\ref{Fig4}. They correspond to local-stress environments that lead to different twin-domain distributions at 20 K under the laser spot. Importantly, the phonon energies change very little among the spots both well above and below $T_\mathrm{s}$, in agreement with our recent finding for BaFe$_2$As$_2$ \cite{Ren2015}. Together with consistent $\Delta_\mathrm{orb}$ and its $T$ dependence reported for twinned and detwinned FeSe \cite{ShimojimaPRB2014,NakayamaPRL2014}, we believe that both the small value of $\Delta_\mathrm{Raman}$ and the departure of $T^*$ from $T_\mathrm{s}$ in FeSe are robust against local stress.

A conservative interpretation of our result is that the Fe-based superconductors exhibit strong nematic magnetoelastic coupling, consistent with recent transport and neutron Larmor diffraction measurements of the 122 family \cite{ManPreprint2015}. The fact that spin interactions appear dominant over orbital interactions in causing the $C_2$ lattice dynamics is consistent with recent inelastic neutron scattering experiments, in which the energy scale of spin anisotropy is found to be greater than that of the orbital ordering in optimally doped BaFe$_{2-x}$Ni$_x$As$_2$ \cite{SongPreprint2015}.

The pronounced magnetoelastic coupling does not prove by itself that the nematic order is driven by magnetism: our data are consistent with the scenario that orbital-driven nematicity lifts the $ab$-degeneracy for the spins and helps stabilize the stripe antiferromagnetic order in the pnictides, which in turn exerts a strong feedback on the lattice that is absent in FeSe (at least above $T^*$). However, it is not unlikely that both spin- and orbital-driven nematicity can only be stabilized in the presence of a deformable lattice, similar to the formation of charge density waves in metals \cite{Gruener1994}. Under such circumstances, the weakness of orbital's influence on the lattice, as demonstrated by the small phonon-energy split in FeSe and the lack of it between $T^*$ and $T_\mathrm{s}$ despite the nearly saturated value of $\Delta_\mathrm{orb}$ at $T^*$ \cite{ShimojimaPRB2014,NakayamaPRL2014}, suggests that orbital interactions alone might not be able to cause the nematic order. In light of recent theoretical proposals for spin-driven nematicity in FeSe without long-range magnetic order \cite{Wang2015,YuPRL2015,Glasbrenner2015,ChubukovPRB2015}, it will be interesting to compare anisotropic spin correlations, either derived from such theories \cite{JiangPRB2009} or in principle measurable by neutron scattering \cite{WangPreprint2015,RahnPRB2015}, to our measured phonon-energy splits, both in the zero-temperature limit and as functions of temperature.

To conclude, we have determined the $E_g$ to $B_{2g} + B_{3g}$ phonon-energy split in FeSe and compared it to those in the Ba(Fe$_{1-x}$Co$_x$)$_2$As$_2$ system. A drastic difference is found both in the much reduced energy split and in the onset of the split in FeSe only below a temperature $T^*$ that is considerably lower than $T_\mathrm{s}$. Our result demonstrates that spin correlations in Fe-based superconductors have a much stronger influence on the lattice than orbital interactions. If the nematic order requires participation of lattice deformation to be fully stabilized, it is unlikely to be driven solely by orbital interactions.

\begin{acknowledgments}

We wish to thank F. Wang and D.-H. Lee for stimulating discussions. Work at Peking University is supported by NSFC (No. 11374024) and MOST (No. 2013CB921903). Work at the IOP, CAS is supported by MOST (Nos. 2011CBA00110 and 2015CB921302), NSFC (Nos. 11374011 and 91221303), and CAS (SPRP-B: XDB07020300). Work at Rice University is supported by the U.S. NSF, No. DMR-1436006 and No. DMR-1308603, and in part by the Robert A. Welch Foundation under Grant No. C-1839.

\end{acknowledgments}

\bibliography{BFA_FeSeRef}

\end{document}